\documentclass[11pt]{article}
\usepackage{amsmath, amssymb, amsfonts}
\usepackage{graphicx}
\usepackage{hyperref}
\usepackage{geometry}
\usepackage{authblk}

\title{Sampling-Based Estimation of Jaccard Containment and Similarity}
\author{Pranav Joshi\thanks{Email: \texttt{pranav.joshi@iitgn.ac.in}}}
\date{July 18, 2025}

\begin{document}

\maketitle

\begin{abstract}
We study the problem of estimating the Jaccard containment $\phi = |A \cap B| / |A|$ between two sets $A$ and $B$, given access only to uniformly sampled subsets $P \subseteq A$ and $Q \subseteq B$. While several sketching-based approaches exist for approximate set similarity, we consider the case where set sizes are known and sketches are not used. We derive an exact likelihood for the observed sample overlap $x = |P \cap Q|$, and show that a binomial approximation—motivated by sparse sampling—is both analytically tractable and empirically accurate in this regime.

We compare this binomial model with a union-based likelihood used in prior work and demonstrate the differences via experiments. We also analyze the posterior distribution of the intersection size $I = |A \cap B|$ under a uniform prior, derive tight upper bounds on the posterior mean squared error (MSE), and provide sample complexity guarantees for achieving desired error and confidence thresholds.

Additionally, we extend our framework to estimate the Jaccard similarity $J = |A \cap B| / |A \cup B|$ using the same sampling model. We derive bounds on the fractional error of $\hat{J}$, showing that it scales as $O(1/\sqrt{x})$, and propose a corrected estimation scheme when MinHash is applied to the sampled subsets. 
These results provide a unified approach to similarity estimation from samples, with implications for large-scale data systems where sketches for full data are expensive or unavailable.
\end{abstract}

\section{Introduction}

Estimating set similarity measures is a fundamental problem in data analysis, with applications in information retrieval, database systems, and streaming algorithms. Among such measures, the \emph{Jaccard containment} of two sets $A, B$—defined as
\[
\phi = \frac{|A \cap B|}{|A|} \in [0,1]
\]
when $A$ is treated as the reference set—is particularly important in asymmetric comparison tasks, such as detecting near-duplicates or containment-based joins.

In large-scale settings, exact computation of $\phi$ may be infeasible, as it requires full knowledge of both sets. Sampling-based estimators that use small random subsets $P \subseteq A$ and $Q \subseteq B$ can be used as scalable alternatives when the sizes $|A|, |B|$ are known, such as in \textit{Oracle} databases.

This paper presents a theoretical analysis of the likelihood models and estimation strategies for Jaccard containment based on random samples, focusing on both empirical performance and statistical guarantees. In particular, we:

\begin{itemize}
    \item Motivate a binomial approximation to the true likelihood under mild assumptions.
    \item Compare this approximation with both an exact combinatorial formulation and a union-based model from prior work.
    \item Use empirical sampling to validate the accuracy of different likelihood models and associated estimators.
    \item Analyze posterior distributions under uniform priors, and derive conditions for bounded error with high probability.
\end{itemize}

While our primary focus is on Jaccard containment, we also address the related and widely-used measure of \emph{Jaccard similarity}, defined as
\[
J = \frac{|A \cap B|}{|A \cup B|}.
\]
Estimating $J$ is essential in applications such as document similarity, deduplication, and clustering, especially when full data access is infeasible. We demonstrate how our sampling-based estimator for intersection size $I = |A \cap B|$ can be used to construct accurate estimates of $J$, and provide theoretical bounds on the error introduced due to sampling. We also propose a correction scheme for MinHash when applied to sampled subsets.

\subsection*{Prior Work}

Numerous approaches have been proposed for estimating set resemblance and containment under various sampling schemes. Notable contributions include:
\begin{enumerate}
    \item {Improved Consistent Weighted Sampling Revisited (I\textsuperscript{2}CWS)}~\cite{bai2017improved}
    \item {SuperMinHash}~\cite{ertl2017superminhash}
    \item {Consistent Sampling with Replacement}~\cite{talwar2006consistent}
    \item {Estimating Set Intersection Using Small Samples}~\cite{henzinger2006intersection}
    \item {On the Resemblance and Containment of Documents}~\cite{broder1997resemblance}
    \item {Selectivity Estimation on Set Containment Search}~\cite{rajaraman2019selectivity}
\end{enumerate}

However, none of these works consider the specific scenario of independent random sampling without sketching, while also assuming that set sizes are known—a setting relevant in systems like Oracle databases.

\clearpage

\subsection*{Contributions}

This paper contributes a structured analysis of sampling-based estimation of Jaccard containment, including:
\begin{itemize}
    \item A formal derivation of a binomial approximation for the likelihood function $P(x \mid I)$.
    \item A comparison between binomial, union-based, and exact models via empirical experiments.
    \item Posterior error bounds for the binomial estimator under a uniform prior.
    \item Analytical expressions for required sample sizes to achieve desired accuracy and confidence.
    \item An extension of the framework to estimate the Jaccard similarity $J = |A \cap B| / |A \cup B|$, including posterior error bounds and practical guarantees on fractional deviation.
    \item A corrected MinHash estimation strategy for sampled data, including batched MinHash, with a combined analysis of sampling and sketching errors.
\end{itemize}

\clearpage
\section{Problem Setup}

\subsection*{Definitions}

Let $A$ and $B$ be two finite sets, with corresponding sample subsets $P \subseteq A$ and $Q \subseteq B$ drawn uniformly at random without replacement. Define:

\begin{itemize}
    \item $N_1 = |A|$, $N_2 = |B|$ — cardinalities of the original sets,
    \item $M_1 = |P|$, $M_2 = |Q|$ — sizes of the sampled subsets,
    \item $I = |A \cap B|$ — true intersection size,
    \item $x = |P \cap Q|$ — observed intersection size in the samples.
\end{itemize}

We are interested in estimating the quantity $I$, or the Jaccard containment $\phi = I / N_1$, using only the observed value $x$ and the known values of $N_1, N_2, M_1, M_2$.

\subsection*{Assumptions}

We operate in the asymptotic regime where the sample sizes are small relative to the sets, yet large enough to permit statistical concentration:

\[
x \le M_1 \ll N_1, \quad x \le M_2 \ll N_2,
\]
\[
1 \ll x \ll I < N_1 \le N_2,
\]
\[
M_1 - x \ll N_1 - I, \quad M_2 - x \ll N_2 - I.
\]

This regime reflects practical scenarios where only small sketches or subsamples of large datasets can be used.

\subsection*{Goal}

Given the observed intersection count $x$, the aim is to estimate the unknown intersection size $I$ (and hence the containment $\phi$), using an appropriate likelihood model $P(x \mid I)$.

We consider three such models:

\begin{enumerate}
    \item A \textbf{binomial approximation}, based on independent sampling assumptions,
    \item A \textbf{union-based model}, used in prior work for sketch-based estimation,
    \item An \textbf{exact combinatorial formulation}, derived from the underlying sampling process.
\end{enumerate}

Each model will be used to derive either a maximum likelihood estimator (MLE) or a moment-based estimator for $I$, and will be compared both theoretically and empirically.

\clearpage
\section{Likelihood Models for \texorpdfstring{$P(x \mid I)$}{P(x | I)}}

We now derive three likelihood models for the observed intersection count $x$, conditioned on the unknown intersection size $I$. These models differ in their assumptions and analytical tractability.

\subsection{Binomial Approximation $P_b(x \mid I)$}

\subsubsection*{Derivation}

Consider the Cartesian product $F = A \times B$. Define a special subset of size $I \subseteq F$, consisting of all pairs $(a, b) \in A \times B$ such that $a = b$ and $a \in A \cap B$. We refer to these as ``matching pairs''.

By sampling $P \subseteq A$ and $Q \subseteq B$, the corresponding set of sample pairs is $P \times Q \subseteq F$. Then, the observed count $x$ equals the number of matching pairs in $P \times Q$, i.e.,
\[
x = |\{(a, b) \in P \times Q \mid a = b\}|.
\]

Assuming each of the $M_1 M_2$ pairs in $P \times Q$ is independently a matching pair with probability $\frac{I}{N_1 N_2}$, we model $x$ as a binomial random variable:
\[
P_b(x \mid I) = \binom{M_1 M_2}{x} \left( \frac{I}{N_1 N_2} \right)^x \left(1 - \frac{I}{N_1 N_2} \right)^{M_1 M_2 - x}.
\]

This approximation ignores dependencies arising from sampling without replacement. It is valid in the regime $M_1 M_2 \ll N_1 N_2$, where the distinction between sampling with and without replacement becomes negligible.

\subsubsection*{Intuition via Grid Visualization}

We may visualize the sample space as a large $N_1 \times N_2$ grid. The $I$ matching pairs lie on the diagonal, with the constraint that no two matches share the same row or column. The sample $P \times Q$ forms a small $M_1 \times M_2$ subgrid, and $x$ is the number of matches observed within this subgrid.

\begin{figure}[h]
    \centering
    \includegraphics[width=0.5\textwidth]{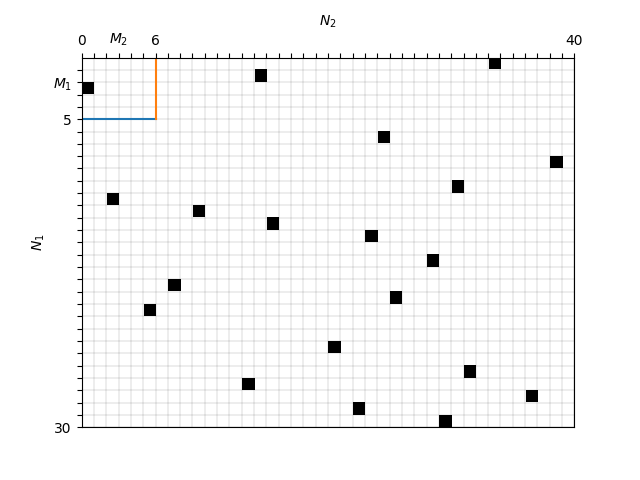}
    \caption{Matching pairs in the $N_1 \times N_2$ grid.}
    \label{fig:grid}
\end{figure}

When the grid is sparse and the subgrid is small, the presence of one match does not significantly influence the others, motivating the use of the binomial approximation.

\subsection{Union-Based Likelihood $P_u(x \mid I)$}

\subsubsection*{Construction}

An alternative model, proposed in prior work on estimating set intersection size, is based on partitioning the universe $A \cup B$ into disjoint components:
\begin{itemize}
    \item $x$ elements from $A \cap B$,
    \item $M_1 - x$ from $A \setminus B$,
    \item $M_2 - x$ from $B \setminus A$.
\end{itemize}

Under this model, the likelihood function is:
\[
P_u(x \mid I) = \frac{\binom{I}{x} \binom{N_1 - I}{M_1 - x} \binom{N_2 - I}{M_2 - x}}{\binom{N_1 + N_2 - I}{M_1 + M_2 - x}}.
\]

\subsubsection*{Limitations}

The union-based model assumes that duplicate elements (i.e., elements appearing in both samples) are treated only once, as in many sketching methods. It does not accurately reflect the true distribution of $x = |P \cap Q|$ when full sample intersection counts are retained. As a result, it introduces significant bias in such settings.

\subsection{Exact Distribution $P(x \mid I)$}

\subsubsection*{Combinatorial Derivation}

We now derive the exact likelihood of observing $x$ matches given the intersection size $I$. We characterize the sample pair $(P, Q)$ by a 5-tuple decomposition based on disjoint components of $A$ and $B$. Let $a$ and $b$ denote the number of additional elements from $B \setminus A$ and $A \setminus B$ included in the samples.

\begin{figure}[h]
    \centering
    \includegraphics[width=0.4\textwidth]{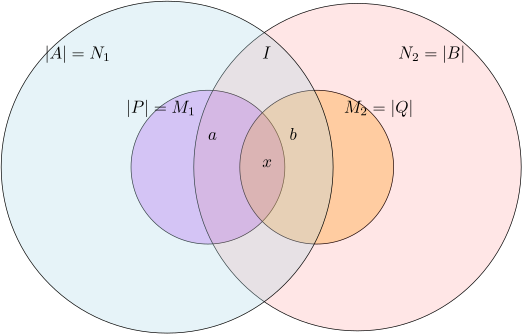}
    \caption{Characterisation of $(P,Q)$ pair as 5-tuple.}
    \label{fig:venn}
\end{figure}

The number of valid configurations for a fixed $a, b, x$ is:

\[
n(a, b, x \mid I) =
\binom{I}{x}
\binom{I - x}{a + b}
\binom{a + b}{a}
\binom{N_1 - I}{M_1 - x - a}
\binom{N_2 - I}{M_2 - x - b}.
\]

Summing over all valid $a$ and $b$, we obtain the total count:
\[
n(x \mid I) = \sum_{a = 0}^{M_1 - x} \sum_{b = 0}^{M_2 - x} n(a, b, x \mid I),
\]

and the corresponding likelihood is:
\[
P(x \mid I) = \frac{n(x \mid I)}{\binom{N_1}{M_1} \binom{N_2}{M_2}}.
\]

\subsubsection*{Intractability}

While this expression is exact, its use in practice is limited by the nested summation and high computational cost, particularly for large $N_1, N_2$. Consequently, we treat it as ground truth for empirical evaluation but do not rely on it for estimator derivation or posterior analysis.

\clearpage
\section{Experimental Comparison of Likelihoods}

We now empirically compare the binomial and union-based likelihood models with the exact distribution, using synthetic data. The goal is to assess how well each model approximates the true distribution of $x$, and how their associated estimators behave in terms of bias and variance.

\subsection*{Experimental Setup}

We generate two synthetic sets $A$ and $B$ of sizes $N_1$ and $N_2$ respectively, with a fixed intersection size $I = |A \cap B|$. The sample subsets $P \subseteq A$ and $Q \subseteq B$ are drawn uniformly at random without replacement, of sizes $M_1$ and $M_2$ respectively.

The observed value $x = |P \cap Q|$ is recorded over multiple trials. This empirical distribution of $x$ is then compared against the binomial and union-based likelihoods.

\subsubsection*{Parameters}

The following values are used throughout the experiments:

\[
N_1 = 10^6, \quad N_2 = 2 \times 10^6, \quad I = 5 \times 10^5, \quad M_1 = 3 \times 10^4, \quad M_2 = 4 \times 10^4.
\]

Each experiment is repeated 100 times to obtain a stable estimate of the empirical distribution.

\subsection*{Distributional Comparison}

The distribution of $x$ generated by sampling is plotted alongside $P_b(x \mid I)$ and $P_u(x \mid I)$. The binomial approximation closely matches the empirical data.

\begin{figure}[h]
    \centering
    \includegraphics[width=0.7\textwidth]{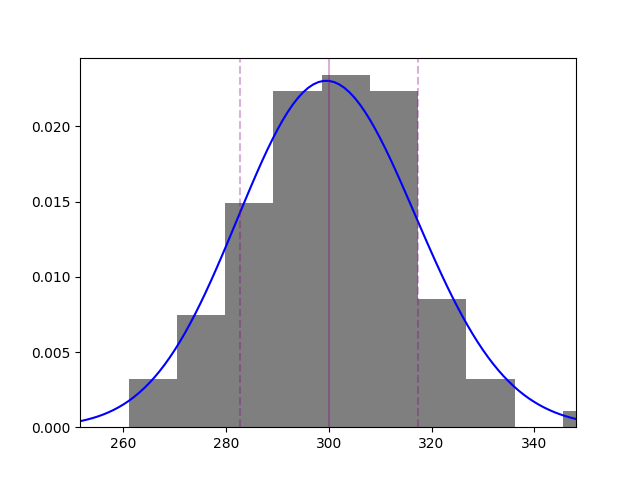}
    \caption{Empirical distribution of $x$ (blue) vs. binomial approximation. Red lines denote $\pm$1 standard deviation.}
    \label{fig:x-zoomed}
\end{figure}

In contrast, the union-based likelihood performs poorly in this setting:

\begin{figure}[h]
    \centering
    \includegraphics[width=0.7\textwidth]{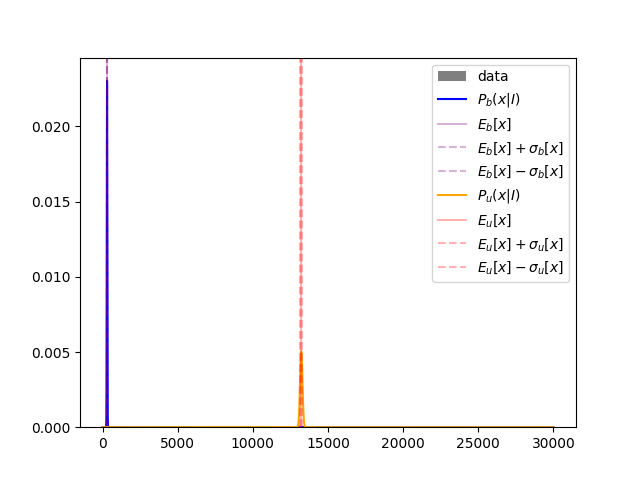}
    \caption{Empirical vs. union-based likelihood distribution for $x$.}
    \label{fig:x-full}
\end{figure}

The computed statistics are:

\[
\begin{aligned}
\mathbb{E}[x] &= 299.84, &\quad \text{Median}(x) &= 300.0, &\quad \text{Std}(x) &= 16.10, \\
\mathbb{E}_b[x] &= 300.0, &\quad \text{Std}_b(x) &= 17.32, \\
\mathbb{E}_u[x] &= 13220.88, &\quad \text{Std}_u(x) &= 79.22.
\end{aligned}
\]

The large discrepancy in mean and variance for $P_u(x \mid I)$ reflects its unsuitability for modeling $x = |P \cap Q|$ directly. 
This is expected, as the union-based model was derived under different assumptions than what we are dealing with.

\subsection*{Estimator Behavior}

We now compare the estimators derived from each likelihood model:

\begin{itemize}
    \item Binomial estimator:
    \[
    \hat I_b = \frac{x}{M_1 M_2} N_1 N_2
    \]
    \item Union-based estimators:
    \[
    \hat I_{u, \text{MLE}}, \quad \hat I_{u, \text{min}} \quad \text{(closed-form expressions from prior work)}
    \]
\end{itemize}

The distribution of these estimators over 100 trials is plotted below.

\begin{figure}[h]
    \centering
    \includegraphics[width=0.7\textwidth]{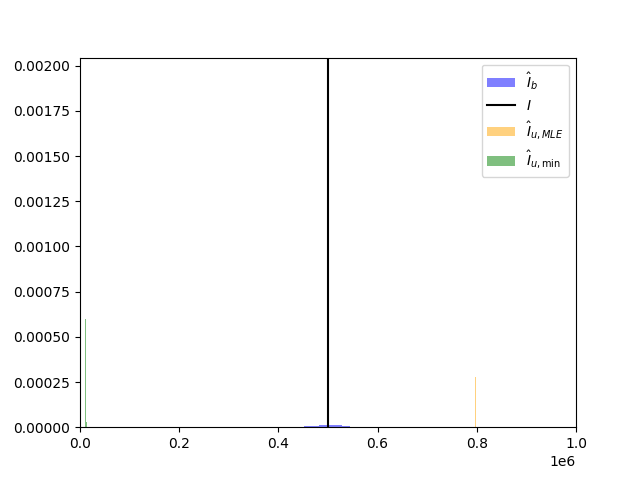}
    \caption{Distribution of estimators: binomial and union-based.}
    \label{fig:estimator-all}
\end{figure}

A zoomed-in plot of $\hat I_b$ shows that it is tightly concentrated around the true value:

\begin{figure}[h]
    \centering
    \includegraphics[width=0.7\textwidth]{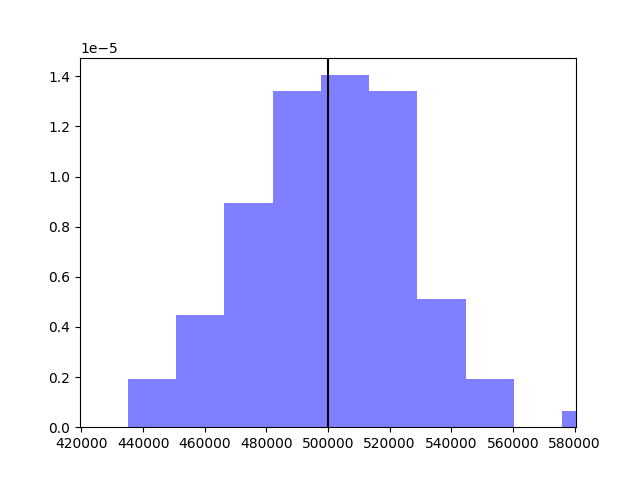}
    \caption{Zoomed-in view of $\hat I_b$ estimator distribution.}
    \label{fig:estimator-binomial}
\end{figure}

These results indicate that the binomial model is not only analytically tractable but also empirically accurate in the considered sampling regime.

\clearpage
\section{Posterior Inference and Error Bounds}

We now analyze the posterior distribution of the intersection size $I$ under the binomial likelihood model and a uniform prior. The goal is to derive guarantees on estimation accuracy, particularly bounding the posterior mean squared error (MSE).

\subsection*{Uniform Prior and Posterior Derivation}

Assume a uniform prior over the interval $[0, N_1]$:
\[
p(I) = 
\begin{cases}
\frac{1}{N_1} & \text{if } I \in [0, N_1], \\
0 & \text{otherwise}.
\end{cases}
\]

Under the binomial likelihood model:
\[
P_b(x \mid I) \propto I^x \left(1 - \frac{I}{N_1 N_2} \right)^{M_1 M_2 - x},
\]
the unnormalized posterior is proportional to a Beta-like function:
\[
B(I) = \frac{1}{N_1 N_2} \cdot \text{Beta}\left(x + 1, M_1 M_2 - x + 1\right)\left( \frac{I}{N_1 N_2} \right).
\]

Let $C(I)$ denote the cumulative distribution function of $B(I)$, and define the normalized posterior as:
\[
p(I \mid x) = \frac{B(I)}{C(N_1)}.
\]

We do not evaluate this posterior directly, but instead derive an upper bound on the posterior MSE of the estimator $\hat I_b = \frac{x}{M_1 M_2} N_1 N_2$.

\subsection*{Bounding Posterior MSE}

The variance of a Beta distribution ~\cite{mood1974statistics, casella2001statistical} with parameters $\alpha = x + 1$ and $\beta = M_1 M_2 - x + 1$ is:
\[
\text{Var}[\theta] = \frac{\alpha \beta}{(\alpha + \beta)^2 (\alpha + \beta + 1)}.
\]

Scaling this by $(N_1 N_2)^2$ gives an upper bound on the MSE of the estimator:
\[
\text{MSE}_{p(I)}(\hat I_b) < \text{Var}_{B}[I] < (N_1 N_2)^2 \cdot \frac{x (M_1 M_2 - x)}{(M_1 M_2)^2 (M_1 M_2 + 1)}.
\]

This implies:
\[
\text{MSE}_{p(I)}(\hat I_b) < \frac{x (N_1 N_2)^2}{(M_1 M_2)^2}.
\]

To express this in terms of the containment $\phi = I / N_1$ or $I / N_2$, we write:
\[
\text{MSE}_{p(\phi_1)}(\hat \phi_1) < \frac{x N_2^2}{(M_1 M_2)^2}, \quad
\text{MSE}_{p(\phi_2)}(\hat \phi_2) < \frac{x N_1^2}{(M_1 M_2)^2}.
\]

If we sample $M_1 = \alpha_1 N_1$ and $M_2 = \alpha_2 N_2$, then the above bound becomes:
\[
\text{MSE}_{p(\phi_1)}(\hat \phi_1) < \frac{1}{N_1 \alpha_1 \alpha_2^2}, \quad
\text{MSE}_{p(\phi_2)}(\hat \phi_2) < \frac{1}{N_2 \alpha_2 \alpha_1^2}.
\]

Suppose we require:
\[
\text{MSE} < \delta^2.
\]
Then it suffices that:
\[
\alpha_1 \alpha_2^2 \ge \frac{1}{N_1 \delta^2}.
\]

For example, if $\delta = 0.01$ and $N_1 = 10^8$, then:
\[
\alpha_2 \ge \sqrt{\frac{1}{N_1 \alpha_1 \delta^2}} = \sqrt{ \frac{1}{10^8 \cdot \alpha_1 \cdot 10^{-4}} }
= \frac{1}{\sqrt{10^4 \alpha_1}}.
\]

If $N_1 \ll N_2$, then maintaining the same $\alpha$ for both may be infeasible due to scaling.

\subsection*{Estimator Validity}

To ensure that the estimator $\hat I_b$ lies in a valid range (i.e., $\hat I_b < N_1$), it is sufficient that:
\[
x < \frac{M_1 M_2}{N_2}.
\]

Let the expected value of $x$ under the binomial model be:
\[
\mu = I \alpha_1 \alpha_2.
\]

We define the failure event $\hat I_b > N_1$ as equivalent to $x > N_1 \alpha_1 \alpha_2$. Let $t = (N_1 - I)\alpha_1 \alpha_2$. Then, using the Chernoff bound ~\cite{boucheron2013concentration, dubhashi2009concentration}:
\[
P(x \ge \mu + t) \le \exp\left( -\frac{t^2}{2(\mu + t/3)} \right) \le \epsilon.
\]

Solving this inequality for $\alpha_1 \alpha_2$ gives a sufficient condition:
\[
\alpha_1 \alpha_2 \ge \frac{\ln(1/\epsilon)}{N_1} \cdot \frac{2 + \frac{\phi}{1 - \phi}}{1 - \phi},
\]
where $\phi = I / N_1$.

For example, with $\phi = 0.9$ and $\epsilon = 10^{-3}$:
\[
\alpha_1 \alpha_2 \ge \frac{3 \ln 10}{N_1} \cdot \frac{11}{0.1} \approx \frac{760}{N_1}.
\]

Thus, sampling with $\alpha = \sqrt{760/N_1}$ ensures the estimator is valid with high probability.

\subsection*{Upper Bound on Error when Estimator is Valid}

Using the constraint $\hat{I}_b < N_1$, one can show that
\[
\text{MSE}_{p(\phi_1)}(\hat{\phi}_1) < \frac{x N_2^2}{(M_1 M_2)^2} < \frac{N_2}{M_1 M_2}
\]
\[
\text{MSE}_{p(\phi_2)}(\hat{\phi}_2) < \frac{x N_1^2}{(M_1 M_2)^2} < \frac{N_1}{M_1 M_2}
\]

We can use these inequalities to ensure that the estimator is accurate, which requires that
\[
\frac{N_2}{M_1 M_2} \le \delta^2
\quad \Rightarrow \quad
\alpha_1 \alpha_2 \ge \frac{1}{\delta^2 N_1}
\]

So, the condition is achieved when:
\[
\alpha_1, \alpha_2 \ge \alpha = \frac{1}{\delta \sqrt{N_1}}
\]

This is an $O(1/\sqrt{N_1})$ quantity as compared to the earlier $O(1/\sqrt[3]{N_1})$ sampling rate we obtained, and is thus a much smaller requirement for achieving the same error threshold $\delta$ under valid estimation.

\clearpage
\section{Sample Complexity for Containment}

We now interpret the analytical bounds derived in the previous section to understand the required sampling rates for accurate and valid estimation of Jaccard containment. In particular, we focus on the relationships between the sample sizes $(M_1, M_2)$, the containment $\phi$, and the desired accuracy and confidence parameters $(\delta, \epsilon)$.

\subsection*{Required Sample Sizes for Error Guarantees}

We now translate the analytical error bounds into concrete requirements on the sample sizes $(M_1, M_2)$ for achieving desired estimation accuracy and confidence.

To ensure that the root mean squared error (RMSE) of the containment estimator $\hat{\phi}_1$ is bounded by a target $\delta$, we first recall the tighter MSE bound derived under the assumption that the estimator is valid:
\[
\text{MSE}_{p(\phi_1)}(\hat{\phi}_1) < \frac{N_2}{M_1 M_2}.
\]
To satisfy $\text{MSE} < \delta^2$, it suffices that:
\[
\alpha_1 \alpha_2 \ge \frac{1}{\delta^2 N_1}, \quad \text{where} \quad \alpha_1 = \frac{M_1}{N_1}, \quad \alpha_2 = \frac{M_2}{N_2}.
\]

In addition to bounding error, we must ensure that the estimator remains valid (i.e., $\hat{I}_b < N_1$). Using a Chernoff bound on the tail of $x$, we showed that the estimator is valid with probability at least $1 - \epsilon$ when:
\[
\alpha_1 \alpha_2 \ge \frac{2 \ln(1/\epsilon)}{(1 - \phi)^2 N_1}.
\]

Taken together, the two conditions imply that to guarantee both accuracy and estimator validity, the sampling rates must satisfy:
\[
\alpha_1 \alpha_2 \ge \max\left\{
    \frac{1}{\delta^2 N_1},\;
    \frac{2 \ln(1/\epsilon)}{(1 - \phi)^2 N_1}
\right\}.
\]

It is worth noting that in our initial posterior analysis—without assuming estimator validity—we derived a weaker sufficient condition:
\[
\alpha_1 \alpha_2 \gtrsim \left( \frac{1}{\delta^2 N_1} \right)^{1/3}.
\]

\subsection*{Tradeoffs Between $M_1$, $M_2$, $\delta$, and $\epsilon$}

Let us assume symmetric sampling, i.e., $\alpha_1 = \alpha_2 = \alpha$. Then the bounds above reduce to:
\[
\alpha \ge \frac{1}{\delta \sqrt{N_1}}, \quad
\alpha \ge \frac{1}{(1 - \phi)} \sqrt{ \frac{2 \ln(1/\epsilon)}{N_1} }.
\]

This implies:
\[
M_1 \ge \frac{ \sqrt{N_1} }{ \delta }, \quad
M_2 \ge \frac{N_2}{N_1} M_1.
\]

These expressions reveal the following tradeoffs:
\begin{itemize}
    \item Stricter accuracy requirements (smaller $\delta$) require larger sample sizes.
    \item Lower containment ($\phi$ small) and very large containment ($(1 - \phi)$ small) also require larger samples.
    \item If $N_2 \gg N_1$, the required $M_2$ can become prohibitively large due to the scaling factor $N_2 / N_1$.
\end{itemize}

In contrast, if we only enforce a looser bound such as:
\[
\text{MSE} < \delta^2 \quad \Rightarrow \quad \alpha \gtrsim \left( \frac{1}{\delta^2 N_1} \right)^{1/3},
\]
we obtain a cubic root dependence, which is weaker and results in lower sample sizes.

\subsection*{Practicality in Real-World Regimes}

Consider a realistic example with:
\[
\phi = 0.1, \quad \delta = 0.01, \quad \epsilon = 0.001, \quad N_1 = 10^8, \quad N_2 = 5 \times 10^8.
\]

Using the tighter condition:
\[
\alpha \ge \frac{1}{\delta \sqrt{N_1}} = \frac{1}{0.01 \cdot \sqrt{10^8}} = \frac{1}{100} = 0.01,
\]

and the validity condition:
\[
\alpha \ge \sqrt{ \frac{\ln(1/\epsilon)}{N_1} \cdot \frac{2}{(1 - \phi)^2} } \approx 0.000403,
\]

we use $\alpha = 0.001$ and obtain:
\[
M_1 = \alpha N_1 = 10^5, \quad
M_2 = \alpha N_2 = 5 \times 10^5.
\]

These values are feasible in many applications such as document similarity, telemetry data, or genome set comparisons. 
However, in extreme-scale problems where $N_2 \gg N_1$, the required $M_2$ can grow quickly, making symmetric sampling rates impractical. 
In such cases, asymmetric sampling rates may be more suitable.

Additionally, one could relax the accuracy requirement (e.g., increase $\delta$ to $0.05$) to substantially reduce the sample sizes.

\subsection*{Sample Complexity Analysis Summary}

The required sample sizes depend polynomially on the inverse accuracy and square root of the reference set size:
\[
M_1 = O\left( \left( \frac{1}{\delta} + \frac{ \sqrt{ \ln(1/\epsilon^2) } }{1 - \phi} \right) \cdot \sqrt{N_1} \right),
\quad
M_2 = O\left( \frac{N_2}{N_1} \cdot M_1 \right).
\]

\clearpage
\section{Estimation of Jaccard Index from Samples}

We now extend our framework to estimate the Jaccard similarity:
\[
J = \frac{|A \cap B|}{|A \cup B|} = \frac{I}{N_1 + N_2 - I}.
\]

Since the sizes $N_1$ and $N_2$ are assumed known and $I$ is estimated via $\hat{I}_b = \frac{x}{M_1 M_2} N_1 N_2$, we define the estimator:
\[
\hat{J} = \frac{\hat{I}_b}{N_1 + N_2 - \hat{I}_b}.
\]

\subsection*{Posterior Error Bounds}

To analyze the error in $\hat{J}$, we begin by recalling that the posterior MSE of $\hat{I}_b$ satisfies:
\[
\text{MSE}_{p(I)}(\hat{I}_b) < \frac{x (N_1 N_2)^2}{(M_1 M_2)^2}.
\]

Let $\hat{J} = \frac{\hat{I}}{N_1 + N_2 - \hat{I}}$ and $J = \frac{I}{N_1 + N_2 - I}$. Then by Taylor expansion and error propagation:
\[
\left| \frac{\Delta \hat{J}}{\hat{J}} \right| \approx \frac{|\Delta \hat{I}|}{|I|} \cdot \left(1 + \frac{I}{N_1 + N_2 - I} \right)
= \frac{|\Delta \hat{I}|}{I} \cdot \left(1 + \frac{N_1}{N_2} \right).
\]

Now since $\text{Var}(\hat{I}) \le \frac{x (N_1 N_2)^2}{(M_1 M_2)^2}$ and $\hat{I}$ concentrates around $I$, we get:
\[
\frac{\Delta \hat{J}}{\hat{J}} \lessapprox \frac{1}{\sqrt{x}} \left(1 + \frac{N_1}{N_2} \right) \le \frac{2}{\sqrt{x}}.
\]

\subsection*{Probability Bounds for $x$}

To ensure that $x \ge c$ with high probability, we apply the lower-tail Chernoff bound. 
Let $\mu = \mathbb{E}[x] = I \alpha_1 \alpha_2$. 
Then for $c < \mu$, we have:
\[
P(x < c) \le \exp\left( -\frac{1}{2} \left(1 - \frac{c}{\mu} \right)^2 \mu \right).
\]

So we solve for the condition:

$$
\exp\left( -\frac{1}{2}(1 - \frac{c}{\mu})^2 \mu \right) \le \epsilon
$$

This will ensure the upper bound of $\epsilon$ on $P(x < c)$. On solving this, we get :

$$
\alpha_1 \alpha_2 \ge \frac{1}{I} ( c + \ln(1/\epsilon) - \sqrt{ \ln(1/\epsilon)(2c + \ln(1/\epsilon)) } ),
$$

In our case, we have $c = 4\delta^{-2}$ where $\delta$ is a upper bound on the standard error in $\hat J$ .

In practice, the $\ln(1/\epsilon)$ term will be small, and can be ignored. Thus, if

$$
\alpha_1\alpha_2 \gtrapprox \frac{4}{\delta^2 I},
$$

then we can claim that the fractional standard error for $\hat J$ will be less that $\delta$ with an arbitarily high confidence (while $\ln(1/\epsilon) << \delta^{-2}$ holds).

\clearpage

\section{Batch-MinHash Estimation of Containment while Sampling}

\subsection{Estimator and Error Bounds}

Suppose from sets $A,B$ we sample $P,Q$ and partition it as 

$$
P = \bigcup_{i=1}^a P_i \quad \text{ and }\quad
Q = \bigcup_{j=1}^b Q_j, 
$$

where

$$
M_1 = |P| = aM, \quad
M_2 = |Q| = bM, \quad
|P_i| = |Q_j| = M. 
$$

Then we can estimate $x_{i,j} = |P_i \cap Q_j|$ using MinHash as $\hat x_{i,j} = \frac{2M\hat J_{i,j}} {\hat J_{i,j} + 1}$. This will result in the RMSE

$$
\text{RMSE}(\hat x_{i,j}|x_{i,j}) \approx \frac{2M}{(J_{i,j} +1)^2}\text{RMSE}(J_{i,j}|J_{i,j}) = \frac{2M\sqrt{J_{i,j}(1-J_{i,j})}}{(1+\hat J_{i,j})^2\sqrt{k}}
$$

Now, we know that :

$$
1+{J_{i,j}} = \frac{2M}{2M-x_{i,j}} $$$$
J_{i,j} = \frac{x_{i,j}}{2M-x_{i,j}} $$$$
1-J_{i,j} = \frac{2M-2x_{i,j}}{2M-x_{i,j}}
$$

Thus

$$
\sqrt{J_{i,j}(1-J_{i,j})} = \frac{\sqrt{2Mx_{i,j}(1-\phi_{i,j})}}{2M-x_{i,j}} $$$$
\implies \frac{\sqrt{J_{i,j}(1-J_{i,j})}}{(1+J_{i,j})^2} = \frac{2M-x_{i,j}}{2M}\sqrt{\frac{1}{2M}x_{i,j}(1-\phi_{i,j})} $$$$
= {(1-\phi_{i,j}/2)}\sqrt{(1-\phi_{i,j})} \sqrt{\frac{x_{i,j}}{2M}} \le \sqrt{\frac{x_{i,j}}{2M}} $$$$
\implies \boxed{\text{MSE}(\hat x_{i,j}|x_{i,j}) \le {\frac{2Mx_{i,j}}{k}}}
$$

We can treat $x_{i,j}$ as i.i.d. random variables with

$$
P(x_{i,j}|\phi) = \text{Bin}(x_{i,j}|\frac{\phi}{N_2}, M^2)
$$
using the Binomial model that we used earlier.

So, we can say that

$$
\text{MSE}[\hat x_{i,j}|\phi] = E_{x_{i,j}}[\text{MSE}[\hat x_{i,j}|x_{i,j}]] \le \frac{2M}{k}E[x_{i,j}|\phi]= \frac{2M^3\phi}{kN_2}
$$

Then, we can say that for $\hat x = \sum_{i,j} \hat x_{i,j}$, we have :

$$
\text{MSE}[\hat x|\phi] < \frac{2abM^3\phi}{kN_2}
$$

Then, for $\hat \phi = \hat x \cdot \frac{N_2}{abM^2}$ , we get : 

$$
\text{MSE}[\hat \phi] = \frac{2N_2\phi}{kabM} = 2\frac{M}{k} \frac{N_2}{M_1M_2}\phi.
$$

\subsection{Unknown distribution for $x_i$}

The estimator that we have constructed, in its full form is

$$
\hat \phi = \frac{2MN_2}{M_1M_2}\sum_{i,j}\frac{\hat J_{i,j}}{1+\hat J_{i,j}}
$$

This estimator has standard error

$$
\text{RMSE}[\hat \phi] < \sqrt{\frac{2M}{k}\frac{N_2\phi}{M_1M_2}} \le \sqrt{\frac{2M}{k}\frac{N_2}{M_1M_2}}
$$

This is assuming that $x_{i,j}$ really does follow a binomial distribution. In case that fails, we can simply use the fact that $x = \sum_{i,j} x_{i,j} = |P\cap Q| < |P| = M_1$ and write :

$$
\text{MSE} [\hat x]  \le \frac{2M \sum _{i,j} x_{i,j}}{k}  = \frac{2Mx}{k} \le  \frac{2MM_1}{k} 
\implies \text{MSE}[\hat \phi] \le \frac{2}{k}\frac{M N_2^2}{M_1M_2^2}  
\implies \text{RMSE}[\hat x] = \frac{N_2}{M_2} \sqrt{\frac{2}{ka}}
$$

\clearpage
\subsection{Unequal Sample Sizes}

Suppose rather than having $|P_i| = |Q_i| = M$ , we have $|P_i| = m_{1,i}$ and $|Q_i| = m_{2,i}$ , one can show, in a similar manner as before, that 

$$
\text{MSE}[\hat x_{i,j}|x_{i,j}] < \frac{m_{1,i} + m_{2,j}}{k}x_{i,j}
$$

This can also be experimentally validated.
For that purpose, using the parameters

$$
N_1 = N_2 = 10000, \quad
I = 5000, \quad
a = b = 30, \quad
\text{and} \quad 
k = 100,
$$

this scatter plot was created by running the experiment 1000 times :

\begin{center}
    \centering
    \includegraphics[height=10cm]{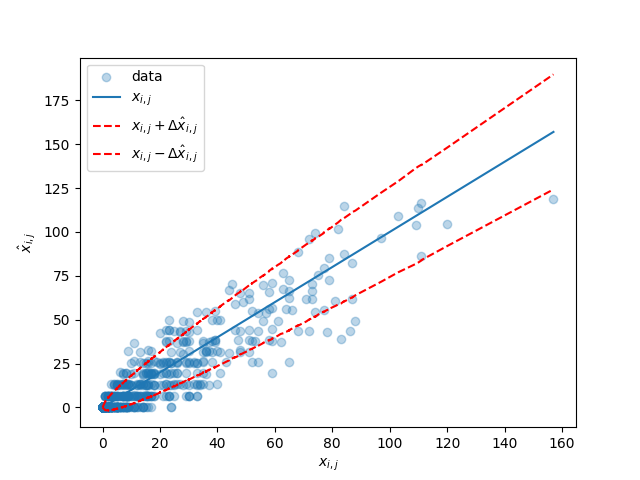}
\end{center}

Here $\Delta \hat x_{i,j}$ is the theoretical RMSE bound that was stated above.

Now suppose, we define 

$$
m_1 = \max_i m_{1,i} 
\quad \text{and} \quad
m_2 = \max_j m_{2,j}
$$

Then, one can prove that

$$
\text{RMSE}[\hat \phi] < \frac{N_2}{M_2}\sqrt{\frac{m_1 + m_2}{kM_1}} < \frac{N_2}{M_2} \sqrt{\frac{2M}{kM_1}}
$$

where $M = \frac{1}{2}(m_1 + m_2)$.

\subsection{Enforcing Independence}

In our derivation, we assumed that $x_i,x_j$ can be treated as independent variables, allowing us
to sum the MSE rather than summing the standard errors. This is a very important step in the derivation, since otherwise, we'll incur an extra factor of $ab$ (using the RMS-AM inequality).
So, we must ensure this always holds. Forturnately, it's easy to do, since all we have to do is make sure that $M_1M_2 << N_1N_2$ . 
This can be done by setting an upper limit of $\alpha_c = 0.1$ on the sampling rates $M_1/N_1$ and $\alpha = M_2/N_2$ , giving $M_1M_2 < 0.01 N_1N_2$ .

With this, our equation for final error becomes

$$
\frac{2}{ak\alpha^2} \le \delta^2 \iff ak \ge \frac{2}{\delta^2 \alpha^2} \ge \frac{200}{\delta^2} 
$$

For a practical scenario, consider $k=200$ and $\delta = 0.1$ , giving us $a \ge 100$ .
The choice of $k=200$ is so that even for standard MinHash over $A,B$ , the standard error in the derived estimatr for Containment would be less than $\sqrt{2/k} = \sqrt{2/200} = 0.1 = \delta$,
making comparison easier.

\subsection{Enforcing Disjointness}

In our derivation, we only used sets. But in practice, $A,B$ are the underlying sets of columns $C_1,C_2$. So, naive batching of columns will most likely lead to non-disjoint batches. 
To prevent this, and at the same time to make sure that variables $x_{i,j}$ do approximate a binomial distribution, we can do stratified sampling.
Use hash function $h_{A}:A \to \{1,2\dots a\}$ and $h_{B}:B \to \{1,2\dots b\}$ to filter from $P,Q$ only those elements that belong to a particular batch.
That is to say, to compute MinHash of batches of $P$ , we perform $a$ passes through $P$ in the database. For batch $P_i$, we will filter the elements to only have those for which $h_A = i$. We will compute the MinHash for all such elements. Similarly, for all other $i\in [a]$. We can stop once $M_1 =|\bigcup_i P_i|$ crosses a given value. 
Since the chosen hash functions $h_A,h_B$ are independent, $x_{i,j}$ will indeed follow the binomial distribution.

Notice that since the original $P$ is irrelevant once $P_i$ is computed, you can sample for $P$ again every call; for example, using the efficient $\text{SAMPLE}$ clause in Oracle for block-sampling. This will also improve the randomness, which is beneficial for our algorithm.
Alternatively, you can use a $\text{GROUP BY}$ operation over $h_A$ over a physical batch and compute parts of MinHash sketches $P_i$ , and combine these parts after fetching to get $P_i$ for all $i\in[a]$ . This will be beneficial for the run-time.

It is very likely that the sample sizes $|P_i|$ and $|Q_j|$ will be close to $M_1/a$ and $M_2/b$, but even if that does not happen, we will get a better estimate and run-time, as will be demonstrated in the next two sub-sections.

\subsection{Runtime Analysis}

Usually, large data is stored in a database. Fetching all the data from the database (even while streaming) to compute MinHash is expensive (in run-time). Thus, it's better to compute MinHash inside the SQL query and fetch only the sketch. This again, faces the issue of an upper limit on the amount of CPU usage per session. So, we must resort to batch processing to compute a MinHash sketch for a batch using the database itself, then fetch the sketches and combine them to get the final sketch.
Since fetching and hashing are the most expensive part of the process, the run-time is usually dominated to their cost in practice.

Moreover, sketches are used when there are many pairs $(A,B)$ for which a metric needs to be evaluated. 
To account for that, we will calculate two run-times $T_1,T_2$,  
defined as the time needed to create a sketch and
to use the sketches to compute the metric, respectively.

For the sake of rigour, let's analyse this using these definitions :

\begin{itemize}
    \item The time needed to fetch a hash value is $F$
    \item The time needed to compute the hash function for a single element using the database itself is $H$.
    \item The time needed to read or write an element from the disk on local device is $S$.
    \item The time needed to perform any elementary operations on data in memory is defined as the unit. Thus, we will use a factor of 1 for it.
\end{itemize}

With these definitions, the run-times of normal MinHash, assuming $M$ batch size is :

$$
T_1(A) \approx kN_1H + k(N_1/M)F + k(N_1/M) + kS
\quad \text{and}
\quad T_2(A,B) \approx 2kS + C_1k
$$

Meanwhile, the run-times for our algorithm is :

$$
T_1'(A) \approx kM_1H + aM_1H + kaF + ka S
\quad \text{and}
\quad T_2'(A,B) \approx kaS + kbS + C_2abk
$$

Here $C_1,C_2$ are small ($<100$) constants.

The $aM_1H$ term is due to the evaluation of $h_A$ for all of $P$, for every $P_i$ . This can be avoided using the $\text{GROUP BY}$ operation discussed earlier.

Based on the timing examples reported on AskTOM (2005) \cite{ora_hash_asktom} and StackOverflow (2019), 
Oracle's internal hash function ORA\_HASH executes in $2 \mu s$ to $10 \mu s$ per call.

And assuming a transmission bandwidth of around $100\text{ Mbps}$ over Oracle SQL*Net (as observed in multiple community benchmarks) , and considering that each hash value is a 32-bit integer,
we can say that $F < 100 \mu s$ as a generous overestimate.
This figure of $100\text{ Mbps}$ is further enforced by the fact that Oracle Net Services documentation (e.g., Oracle Database 21c) \cite{oracle-optimizing-network} uses $100 \text{ Mbps}$ bandwidth as a design example for buffer tuning, indicating this as a typical LAN throughput scenario.
But as demonstrated in Marchwicki (2015) \cite{jdbc-fetch-performance-blog}, when using a small JDBC fetch size (10 rows), $F$ can degrade to (an overestimate of) $100\mu s$ due to protocol overhead and high round-trip latency. This represents a worst-case scenario for bulk data retrieval.

Even with the overestimate, in practice, $F/M < H$ and $aF < M_1H$, as will be evident in the examples in the next two sub-sections, making the hashing the main bottle-neck. 
Thus, our algorithm runs faster as we have at least 5 times lesser cost of hashing than the standard MinHash in terms of $T_1'$ and $T_1$. This is assuming $M_1 < 0.1 N_1$ was enforced as per the previous sub-section.

As for the evaluation of $\phi(A,B)$ for $n^2$ ordered pairs $(A,B)$ after the sketches are obtained, it's better to reduce $b$ before-hand by merging many of the sketches for a particular $B$ . This works since the standard error guarantees are independent of $b$ . 
This can drastically reduces the full run-time $\sum_{(A,B)} T_2'[A,B]$ from $O(n^2abk)$ to $O(n^2ab_\text{red}k)$ .

\clearpage

\subsection{Experimental Validation}

We run the algorithm 30 times, with the values :

$$
N_1 = N_2=  10^6,\quad
I = 5\times10^5,\quad
a = 100,\quad
b = 10,\quad
k = 200,\quad
M_1 = M_2 = 10^5.
$$
    
Each iteration, the seeds for the hash functions for MinHash and $h_A,h_B$ are chosen randomly at the start of the algorithm from the range of $1-2^{32}$ to $2^{32}$ ensuring minimal collisions.
The resultant distributions are :

\includegraphics[height=10cm]{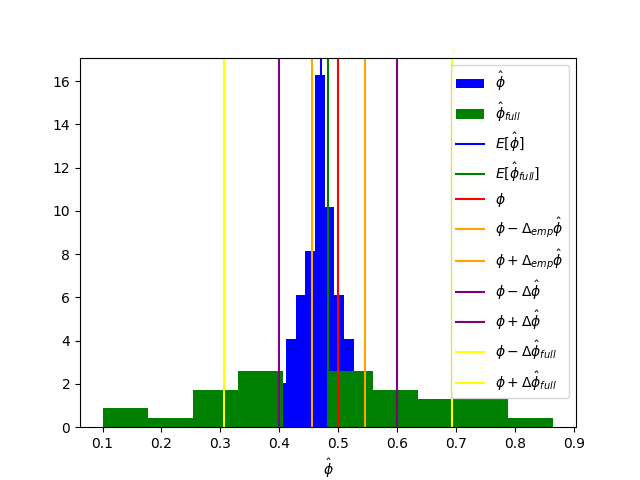}

Here $\hat\phi_{full}$ (green distribution) is the estimate one would get by combining the MinHash sketches for the batches to obtain the sketches $\text{MinHash}[P],\text{MinHash}[Q]$ and estimating $\hat J_{P,Q}$ and from that $\hat x$. 

For the estimate $\hat \phi$ computed by our algorithm, the empirically found standard error $\Delta_{emp}\hat \phi$ is less than the theoretical upper bound $\Delta \hat \phi$.
In comparison, running standard MinHash on $A,B$ and using the estimate $\hat J$ to compute $\hat \phi$ would give the standard error of around $\sqrt{2/k}$ (not an upper bound), which would be 0.1 for $k=200$. 
As we can see from the graph, our algorithm achieves better (or comparable) performance in much less time and moderate space requirements.

It is also worth noting from the figure that the estimator has a small negative bias. Even with theoretical bias corrections of up to second order (using Taylor series expansion), not much success was found in removing it.
The reason for the bias remains unclear. But despite the bias, since the RMSE is small, the estimator performs well.

\clearpage

\section{Sketch of samples to compute Jaccard Index}

\subsection{Estimator and Error Bounds}

Suppose the Jaccard index $J'$ of the samples $P,Q$ is estimated as $\hat J'$ using MinHash, with knowledge of $M_1,M_2,N_1,N_2$ , then since we can write

$$
J' = \frac{x}{M_1 + M_2 - x} 
$$
$$
\implies x = \frac{M_1+M_2}{(1/J') + 1} 
$$
$$
\implies \hat J  = \frac{M_1+M_2}{(1+\frac{1}{J'})(\frac{1}{N_1} + \frac{1}{N_2})M_1M_2 - M_1-M_2} 
$$
$$
= \frac{J' (M_1^{-1} + M_2^{-1})}{(N_1^{-1}+N_2^{-1}) - J'((M_1^{-1}+M_2^{-1}) - (N_1^{-1}+N_2^{-1}))} 
$$
$$
= \frac{J'}{r - J'(1-r)}
$$

where 

$$
r = \frac{N_1^{-1} + N_2^{-1}}{M_1^{-1}+M_2^{-1}} ,
$$

we know that 

$$
\frac{\Delta_s \hat J}{\hat J} \le \frac{2}{\sqrt x} = 2 \frac{\sqrt{1+J'^{-1}}}{\sqrt{M_1+M_2}} 
$$
$$
\implies \Delta_s \hat J  \le 2\frac{\hat J}{J'} \frac{\sqrt{J'(1+J'^{-1})}}{\sqrt{M_1+M_2}} = \frac{\sqrt{3}\hat J / J'}{\sqrt{M_1+M_2}} 
$$
$$
\sqrt{\frac{3}{(M_1+M_2)(r-J'(1-r))^2}}
$$

This is the error due to sampling. But there is another source of error, the MinHash itself.
Suppose the (absolute) error in $J'$ due to MinHash is $\delta_m$ , then, the error propogates to $\hat J$ as 

$$
\Delta_m \hat J \approx [\frac{1}{r - J'(1-r)} + \frac{(1-r)}{(r-J'(1-r))^2}]\delta_m \\ = \frac{1 - J'(1-r)}{(r-J'(1-r))^2}\delta_m
$$

Thus, the full error is

$$
\Delta \hat J \lessapprox \sqrt {[\frac{(1-J'(1-r))}{(r-J'(1-r))^2}\delta_m]^2 + \frac{3}{(M_1+M_2)(r-J'(1-r))^2}}
$$

\subsection{Experimental Validation}

Just like before, let's have $A = \{-N_1+I+1\dots I\}$ and $B = \{1 \dots N_2\}$ . From this, we'll sample $P,Q$ randomly and compute the MinHash sketches for them using some value of $k$ . This will give us the distribution of $\hat J$ . From this, we can get the empirical bias and standard error in the estimator's distribution . But that's not enough since our estimates of standard error for the posterior are now dependent on the estimator itself. Thus, it's better to compute the Z score $Z_i = \frac{\hat J - J}{\Delta \hat J}$ for each run of the experiment and then look at the emperical distribution of the Z-scores. It should be a distribution centered around 0 with low standard deviation.

Running the experiments for 

$$
N_1 = 10^5, \newline
N_2 = 2\times 10^5, \newline
I = 8\times 10^4, \newline
M_1 = M_2 = 2\times 10^4, \newline
\text{and } k = 1000,
$$

we get these results :

\begin{figure}[h]
    \includegraphics[width=0.5\textwidth]{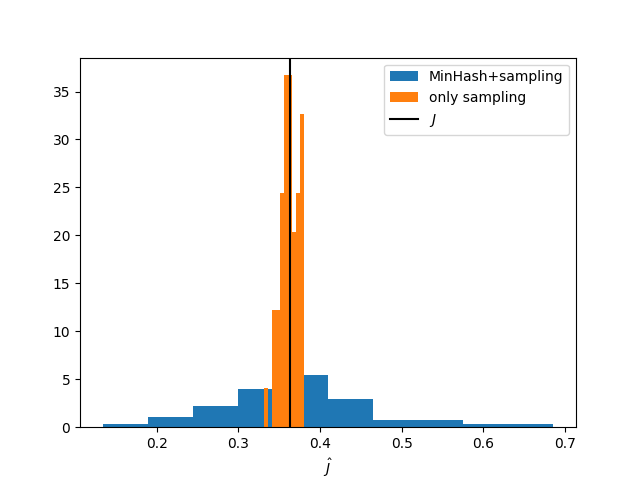}
    \includegraphics[width=0.5\textwidth]{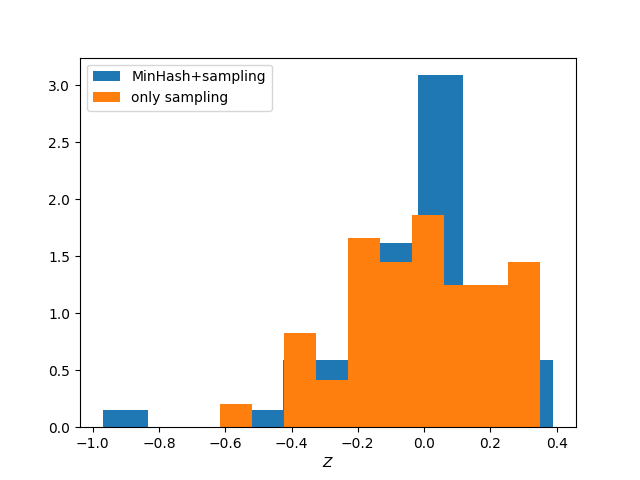}
    \caption{Distribution of $\hat J$ and its Z-score}
    \label{fig:J}
\end{figure}

\clearpage

\section{Batch-MinHash Estimation of Similarity while Sampling}

\subsection{Estimator and Error Bounds}

Suppose MinHash sketches of columns with underlying sets $A,B$ are computed for $a$ batches $A_i \subset A$ and $b$ batches $B_i \subset B$ , 
with every batch being of size $M$ , then we consider $J'_{i,j}$ as the estimated Jaccard index of the pair $(A_i,B_j)$ from the MinHash sketches, 
and $\hat J_{i,j}$ as the estimate for the Jaccard index $J$ for the pair $(A,B)$ , using the value $J'_{i,j}$ . 
Then, for the estimator

$$
\hat J = \frac{1}{ab}\sum_{i,j} \hat J_{i,j}
$$

the standard errors due to sampling and MinHash are :

$$
\Delta_s \hat J  \le \sqrt{\frac{3}{(ab)^2}\sum_{i,j} {(r(1+J'_{i,j})-J'_{i,j})^{-2}}(2M)^{-1}}
$$
$$
\Delta_m \hat J \le \sqrt{\frac{1}{(ab)^2}\sum_{i,j} \frac{(1-J'(1-r))^2}{(r(1+J'_{i,j})-J'_{i,j})^{4}}\delta_{m,i,j}^{2}}
$$

This is assuming that all pairs give valid estimates. Of course, some will not, in which case, we'll remove them from the root-mean-square calculation and use the actual number of valid pairs, rather than $ab$ .

Note that here, by "valid", I mean $\hat I_{i,j} \le N_1$ .

Since for MinHash with $k$ hash functions, you have the frequentist standard deviation

$$
\delta _{m.i,j} \approx \sqrt{\frac{J'_{i,j}(1-J')_{i,j}}{k}},
$$

so we can write

$$
\Delta_m \hat J \le 
\frac{1}{\sqrt k}\sqrt{\frac{1}{(ab)^2}\sum_{i,j} \frac{(1-J'(1-r))^2}{(r(1+J'_{i,j})-J'_{i,j})^{4}}J'_{i,j}(1-J'_{i,j})}
$$

\clearpage

\subsection{Removal of Terms}

If $V$ is initialised to the set of pairs $(i,j)$ that give valid estimates $J_{i,j}'$ , then we get 

$$
\Delta_m \hat J \le {\cal M} = \frac{1}{\sqrt k}\sqrt{\frac{1}{|V|^2}\sum_{(i,j)\in V} \frac{(1-J'(1-r))^2}{(r(1+J'_{i,j})-J'_{i,j})^{4}}J'_{i,j}(1-J'_{i,j})}
$$

$$
\Delta_s \hat J \le {\cal S} = \sqrt{\frac{3}{|V|^2}\sum_{(i,j)\in V} {(r(1+J'_{i,j})-J'_{i,j})^{-2}}(2M)^{-1}}
$$

Now consider the removal of a high error pair $(i,j)$ from $V$ . We want to know whether or not this will decrease the standard deviation.

We can use this approximation :

$$
\Delta_{i,j} {\cal M}^2 \approx \frac{2}{|V|}{\cal M}^2 
- 
\frac{1}{k|V|^2}\frac{(1-J'(1-r))^2}{(r(1+J'_{i,j})-J'_{i,j})^{4}}J'_{i,j}(1-J'_{i,j})
$$

Similarly, 

$$
\Delta_{i,j}{\cal S}^2 = \frac{2}{|V|}{\cal S}^2 - \frac{3}{2M|V|^2}(r(1+J'_{i,j})-J'_{i,j})^{-2}
$$

Now, we will drop any pair that has $\Delta_{i,j} {\cal M}^2 + \Delta_{i,j}S^2 < 0$ . That is to say,

$$
\frac{1}{k}\frac{(1-J'(1-r))^2}{(r(1+J'_{i,j})-J'_{i,j})^{4}}J'_{i,j}(1-J'_{i,j}) + 
\frac{3}{2M}(r(1+J'_{i,j})-J'_{i,j})^{-2}
$$
$$
> 2 |V| ({\cal M}^2 + {\cal S}^2) 
$$

We can keep on removing pairs and recomputing $\cal M,S$ until no such pair exists that satisfies this condition. For the best effect, we should only care about the term with the maximum value of 

$$
C_{i,j} = C_{i,j,m} + C_{i,j,s}
$$

where

$$
C_{i,j,m} = \frac{1}{k}\frac{(1-J'(1-r))^2}{(r(1+J'_{i,j})-J'_{i,j})^{4}}J'_{i,j}(1-J'_{i,j})
$$
$$
C_{i,j,s} = \frac{3}{2M}(r(1+J'_{i,j})-J'_{i,j})^{-2}
$$

In case our standard error is still worse than what it would be if we had used the MinHash sketches for the batches to create the sketches for $A,B$ , then we'll simply use the full MinHash estimate $\hat J_{fm}$ for the Jaccard index.

\subsection{Removal based Estimation Algorithm}

\noindent
\begin{align*}
&\textbf{for } i \in \{1\dots a\} \textbf{ :} \\
&\quad\quad \textbf{for } j \in \{1\dots b\} \textbf{ :} \\
&\quad\quad\quad\quad J'_{i,j} := |\text{MinHash}[A_i] \cap \text{MinHash}[B_j]|/k \\
&\text{MinHash}[A] = [\min_i\text{MinHash}[A_i,t]]_t\\
&\text{MinHash}[B] = [\min_j\text{MinHash}[B_j,t]]_t\\
&\hat J_{fm} := |\text{MinHash}[A] \cap \text{MinHash}[B]|/k \\
&V := \left\{(i,j) \mid \hat I(J'_{i,j}, M, N_1, N_2) \le N_1 \right\} \\
&\textbf{for } (i,j) \in V \textbf{ :} \\
&\quad\quad \text{Compute } C_{i,j} \\
&\textbf{while } |V| > 1 \textbf{ :} \\
&\quad\quad (i,j) := \arg\max_{(i,j) \in V} C_{i,j} \\
&\quad\quad \text{Compute } {\cal M}^2(V),\; {\cal S}^2(V) \\
&\quad\quad \textbf{if } C_{i,j} >  2 |V| ({\cal M}^2 + {\cal S}^2) \textbf{ :} \\
&\quad\quad\quad\quad V := V \setminus \{(i,j)\} \\
&\quad\quad\quad\quad \textbf{continue} \\
&\quad\quad \hat J := \frac{1}{|V|} \sum_{(i,j)\in V} \hat J_{i,j} \\
&\quad\quad \textbf{break} \\
&\textbf{if } (|V| = 1) \textbf{ or } ({\cal M}^2 + {\cal S}^2 < \frac{1}{k} \hat J_{fm}(1 - \hat J_{fm})) \textbf{ :} \\
&\quad\quad \hat J := \hat J_{fm}
\end{align*}

\clearpage
\section{Conclusion}

\subsection*{Summary of Insights}

This paper presented a theoretical and empirical investigation into estimating Jaccard containment via random sampling. The main insights are:

\begin{itemize}
    \item The exact likelihood $P(x \mid I)$ can be derived combinatorially, but is computationally intractable for large sets.
    \item A binomial approximation, motivated by sparse sampling, provides a tractable and accurate model in the relevant regime.
    \item The commonly used union-based model performs poorly when applied directly to sample intersection counts.
    \item Empirical evaluations confirm that the binomial model supports a low-bias, low-variance estimator $\hat I_b$.
    \item Posterior analysis under a uniform prior yields tight MSE bounds and clear sampling rate requirements to ensure estimator accuracy and validity.
\end{itemize}

\subsection*{Limitations}

While the binomial model is effective under the assumed conditions ($M_1 M_2 \ll N_1 N_2$), its accuracy may degrade outside this regime. Additional limitations include:

\begin{itemize}
    \item The analysis assumes known values of $N_1$ and $N_2$, which may not hold in all settings.
    \item Practical systems may introduce runtime or memory constraints that this work does not model.
    \item The assumption of uniformly random sampling without replacement may not reflect biased or hash-based selection strategies.
\end{itemize}

\subsection*{Future Directions}

Several extensions and refinements are worth exploring:

\begin{itemize}
    \item Incorporating runtime constraints and memory limits from real database or analytics pipelines.
    \item Extending the analysis to non-uniform priors and other Bayesian formulations.
    \item Developing improved approximations beyond the binomial model, especially in the dense-sample regime.
    \item Generalizing the framework to estimate other asymmetric measures such as conditional containment or weighted intersection.
\end{itemize}

\subsection*{Code and Reproducibility}

All code used for experiments, empirical validation, and figure generation is available at:

\begin{center}
\url{https://github.com/pranav-joshi-iitgn/JCE}
\end{center}


\begin{thebibliography}{99}

\bibitem{bai2017improved}
K. Bai, C. Li, and Y. Li.
\newblock Improved Consistent Weighted Sampling Revisited (I\textsuperscript{2}CWS).
\newblock \textit{arXiv preprint arXiv:1706.01172}, 2017.

\bibitem{ertl2017superminhash}
C. Ertl.
\newblock SuperMinHash: A New Minwise Hashing Algorithm for Jaccard Similarity Estimation.
\newblock \textit{arXiv preprint arXiv:1706.05698}, 2017.

\bibitem{talwar2006consistent}
K. Talwar and D. Panigrahy.
\newblock Consistent Sampling with Replacement.
\newblock Technical Report, 2006. Available at \url{http://kunaltalwar.org/papers/wtdjacc.pdf}

\bibitem{henzinger2006intersection}
M. Henzinger.
\newblock Estimating Set Intersection Using Small Samples.
\newblock In \textit{Random Structures and Algorithms}, 2006.

\bibitem{broder1997resemblance}
A. Z. Broder, S. C. Glassman, M. S. Manasse, and G. Zweig.
\newblock On the Resemblance and Containment of Documents.
\newblock In \textit{Proc. of the 1997 IEEE SEQUOIA Workshop}, 1997.

\bibitem{rajaraman2019selectivity}
A. Rajaraman and J. D. Ullman.
\newblock Selectivity Estimation on Set Containment Search.
\newblock In \textit{Mining of Massive Datasets}, Springer. \url{https://link.springer.com/article/10.1007/s41019-019-00104-1}

\bibitem{hoeffding1963probability}
W. Hoeffding.
\newblock Probability Inequalities for Sums of Bounded Random Variables.
\newblock \textit{Journal of the American Statistical Association}, 58(301):13–30, 1963.

\bibitem{boucheron2013concentration}
S. Boucheron, G. Lugosi, and P. Massart.
\newblock \textit{Concentration Inequalities: A Nonasymptotic Theory of Independence}.
\newblock Oxford University Press, 2013.

\bibitem{mood1974statistics}
A. M. Mood, F. A. Graybill, and D. C. Boes.
\newblock \textit{Introduction to the Theory of Statistics}.
\newblock McGraw-Hill, 3rd edition, 1974.

\bibitem{casella2001statistical}
G. Casella and R. L. Berger.
\newblock \textit{Statistical Inference}.
\newblock Duxbury Press, 2nd edition, 2001.

\bibitem{dubhashi2009concentration}
D. P. Dubhashi and A. Panconesi.
\newblock \textit{Concentration of Measure for the Analysis of Randomized Algorithms}.
\newblock Cambridge University Press, 2009.

\bibitem{ora_hash_asktom}
Oracle Corporation,
\textit{Ask TOM: ORA\_HASH Function Performance},
2005. Available at:
\url{https://asktom.oracle.com/ords/f?p=100:11:0::NO::P11_QUESTION_ID:26043356526642}. Accessed: July 20, 2025.


\bibitem{oracle-optimizing-network}
Oracle Corporation,
\textit{Oracle Net Services Administrator's Guide - Optimizing Performance (Example Using 100 Mbps Link)},
Oracle Database 21c,
2021. Available at:
\url{https://docs.oracle.com/en/database/oracle/oracle-database/21/netag/optimizing-performance.html}.
Accessed: July 20, 2025.

\bibitem{jdbc-fetch-performance-blog}
Jacek Marchwicki,
\textit{JDBC Fetch-Size Performance},
Make Java Faster Blog, June 2015. Available at:
\url{https://makejavafaster.blogspot.com/2015/06/jdbc-fetch-size-performance.html}.
Accessed: July 20, 2025.

\end{thebibliography}
\end{document}